\documentclass[onecolumn,showpacs,preprintnumbers,amsmath,amssymb,10pt]{revtex4}

\usepackage{graphicx}
\usepackage{dcolumn}
\usepackage{bm}
\newcommand{\be}{\begin{equation}}
\newcommand{\ee}{\end{equation}}
\newcommand{\bd}{\begin{displaymath}}
\newcommand{\ed}{\end{displaymath}}
\newcommand{\BE}{\begin{eqnarray}}
\newcommand{\EE}{\end{eqnarray}}

\begin{document}


\title{Effects of noise and confidence thresholds \\
in nominal and metric Axelrod dynamics of social influence}

\author{Luca De Sanctis} 
\email{lde_sanc@ictp.it}
\author{Tobias Galla }%
\email{galla@ictp.it}
\affiliation{The Abdus Salam International Centre for Theoretical Physics, 
Strada Costiera 11, 34014 Trieste, Italy}

\date{\today}

\begin{abstract}
We study the effects of bounded confidence thresholds and of
interaction and external noise on Axelrod's model of social
influence. Our study is based on a combination of numerical
simulations and an integration of the mean-field Master equation
describing the system in the thermodynamic limit. We find that
interaction thresholds affect the system only quantitatively, but that
they do not alter the basic phase structure. The known crossover
between an ordered and a disordered state in finite systems subject to
external noise persists in models with general confidence
threshold. Interaction noise here facilitates the dynamics and reduces
relaxation times. We also study Axelrod systems with metric features,
and point out similarities and differences compared to models with
nominal features. Metric features are used to demonstrate that a small
group of extremists can have a significant impact on the opinion
dynamics of a population of Axelrod agents.
\end{abstract}

\pacs{Valid PACS appear here}
\maketitle

\
\section{Introduction}
Given the increasing influence of mass media, globalisation,
electronic communication and intercontinental travel the apparent
persistence of cultural diversity seems surprising. To study this
problem Axelrod \cite{Axelrod} proposed a simple agent-based model
to study how cultural features disseminate. In particular the model
addresses the question of how cultural diversity can result from
locally attractive interaction. Axelrod's model is -in the language of
physics and dynamical systems theory- a cellular automaton with a set
of discrete degrees of freedom placed on a discrete spatial lattice,
updated in time through specific interaction rules, and is easily
simulated computationally.  The simulation and theoretical analysis of
social and economic systems has in the recent years been taken up by
the statistical physics community (see
e.g. \cite{Book1,Book2,Book3,Book4,Book5} and references therein).
Social systems are here interpreted as many-particle problems and
tools originally developed to study physical systems have been
transferred and adapted to the study of socio-economic models in
different contexts. See also \cite{flachebook} for a sociological
perspective of agent-based modelling.

One of the first systematic studies of the Axelrod model is the one of
\cite{Castellano}, where a phase between a monocultural state
(referred to as globalisation) and a multi-cultural phase (referred to
as polarisation) has been identified as a function of the degree of
variation of the initial conditions chosen before running the Axelrod
automaton. In terms of physics this transition is one between ordered
and disordered states. Axelrod's two-body interaction is attractive as
such, and drives the system to order. Two particles can only interact,
however, if their states are not fully distinct from each other. These
interaction barriers lead to potential jamming. The model thus
exhibits some similarity with kinetically constrained models of glass
forming materials \cite{ritortsollich, garrahan}. If the disorder of
the initial conditions is higher than some critical value, the kinetic
arrest occurs after a finite number of time-steps, and the systems
remains jammed in a disordered state. For low enough initial disorder
in the random initial conditions, however, the ordering dynamics
continues indefinitely in infinite systems. In finite systems full
convergence is reached at finite times, and the dynamics comes to a
halt when any remaining disorder has been eliminated, resulting in a
fully monocultural `globalised' state.

A variety of extensions and variations of Axelrod's original model
have been proposed, Klemm et al. have for example studied the
effects of noise on the jamming behaviour of Axelrod systems 
\cite{Klemmnoise, Klemm2},
furthermore the effects of mass media influence has recently been
addressed in \cite{GCT,Gonzales}. The
model, originally defined on a square lattice, has furthermore been
simulated on a variety of complex networks in order to study the
effects of the topology of the underlying web of interactions on the
dynamics and convergence properties \cite{Maxi3}.

One of the obvious shortcomings of Axelrod's original model and the
subsequent variations is the fact that a metric structure is missing
in opinion space. Opinions $\sigma_{if}$ of agent $i$ on feature $f$
take values in a discrete set, usually labelled by $\{1,.\dots,q\}$,
and agents are classified only as to whether they have the same
opinion on a certain issue ($\sigma_{if}=\sigma_{jf}$), or whether
their opinions are different ($\sigma_{if}\neq\sigma_{jf}$). No notion
of partial agreement on a feature is present, agents can either fully
agree on a certain issue $f$ or they disagree (a more formal
definition of the model will follow below). This assumption has been
relaxed in a sociological context for example in
\cite{Flache1a,Flache1,Flache2}, and the notion of metric spaces has been
introduced. For such features a gradual distinction of a degree of
agreement can be defined, for example given by
$|\sigma_{if}-\sigma_{jf}|^m$, where the integer $m$ is referred to as
`moderation' \cite{Flache1a}.

As a second drawback, no notion of different tolerance levels or
inclination to change one's opinion is present in the original
setup. Interaction between agents in Axelrod's original model can
occur once they agree on at least one out of a number $F$ of
features. $F$ along with $q$ mentioned above are the main model
parameters in Axelrod's original setup. From a sociological point of
view it is interesting to study the model in more generality, and to
introduce a `confidence threshold' $\vartheta\in\{0,\dots,F\}$, so
that agents have the potential to interact only if they agree on
(strictly) more than $\vartheta$ out of $F$ issues. This is referred to
as `bounded confidence' in the sociological literature, Axelrod's
original model corresponds to minimal confidence threshold
($\vartheta=0$), in which interaction is possible rather easily. Large
values of the threshold $\vartheta$ systematically suppress the
potential to interact, corresponding to more and more conservative
agents, who do not change opinion easily.

Studies of different models addressing either of these two points
e.g. through the introduction of continuous opinion states can be
found in
\cite{Redner2,Redner3,Weisbuch1,Weisbuch1a,Weisbuch2,Weisbuch3,Stauffer}. A
first analysis of the effects of metric features and confidence
thresholds in the context of the Axelrod model has been conducted in
\cite{Flache1,Flache2}. These studies focus mostly on numerical
simulations and is mostly restricted to a specific choices of the model
parameters $F$ and $q$. The aim of the current work is to complement
and extend the analysis of
\cite{Flache1,Flache2} through a more general study of the model in parameter space. We also provide analytical results based on a Master
equation approach \cite{Castellano} for the Axelrod model with nominal
features and general confidence threshold. In addition, the
introduction of a metric model allows us to address issues such as
extremism in the context of Axelrod opinion dynamics, which are not
captured by the conventional `nominal' formulation.

\section{The Model}
The system is composed of $N=L\times L$ agents fixed on the nodes of
square lattice of lateral extension $L$. For simplicity we consider
periodic boundary conditions in both spatial dimensions. The state of
agent $i\in \{1,\dots,N\}$ is characterised by an opinion vector
$\vec{\sigma}_i=(\sigma_{i1},\dots,\sigma_{iF})$, where the integer
$F>1$ denotes the number of cultural `features' in the model. Each
component $\sigma_{if}$ then indicates the opinion of agent $i$ on
issue $f$. In Axelrod's original formulation each component
$\sigma_{if}$ takes one of the $q$ values $\{1,\dots,q\}$ at each time
step, so that each spin $\vec{\sigma}_i$ describes one of $q^F$
cultures. Initially each $\sigma_{if}$ is drawn at random from the set
$\{1,\dots,q\}$ with no correlations between agents or features. The
model parameter $q$ hence measures the degree of disorder in the
random initial spin configuration.

We will in the following distinguish between `nominal' and `metric'
features as suggested in \cite{Flache1,Flache2}. We first describe the
dynamics of the nominal Axelrod model. Here, the system evolves in
time by iteration of the following steps:

\begin{enumerate}
\item\label{uno} Select one spin $i\in\{1,\dots,N\}$ at random. 
Subsequently select one of its four nearest neighbours at random. 
Call this second spin $j$.
\item Compute the overlap 
$\omega(i,j)=\sum_{f=1}^F\delta_{\sigma_{if},\sigma_{jf}}\in\{0,\dots,F\}$ 
between spins $i$ and $j$ (with $\delta_{\sigma,\sigma'}$ the Kronecker delta).
\item If $\omega(i,j)=F$ continue with (v) 
(spins $\vec{\sigma}_i$ and $\vec{\sigma}_j$ are in identical states).
\item Set the probability for $i$ and $j$ to interact to 
$I=\delta$ if $\omega(i,j)\leq \vartheta$ and to 
$I=\omega(i,j)/F$ if $\omega(i,j)>\vartheta$. 
Then with probability $1-I$ leave spins $i$ and $j$ unchanged. 
With probability $I$ spins $i$ and $j$ perform the following interaction: 
choose one feature $f$ at random such that $\sigma_{if}\neq\sigma_{jf}$. 
Such a feature exists as $\omega(i,j)<F$. 
Then set $\sigma_{if}=\sigma_{jf}$. 
\item\label{cinque} External noise. With probability 
$\gamma$ perform the following: choose 
one spin $i$ and one feature $f$ at random. 
Set $\sigma_{if}$ to a value chosen randomly from 
$\{1,\dots,q\}$.
\item\label{sei} Resume at (i). 
\end{enumerate}

We will refer to one cycle (i)-(vi) as a {\em microscopic} time-step
in the following. At system size $N$ the duration of such a step is
taken to be $\Delta t=1/N$. In general we will present the
time-evolution of the system mostly in terms of {\em macroscopic} time
units $t$, so that one unit of time $t$ corresponds to $N$ microscopic
interaction cycles, i.e. on average to one (attempted) update per
spin.

In the above dynamics $\vartheta$ is the interaction threshold
mentioned above.  In the absence of interaction noise ($\delta=0$)
neighbouring agents have the potential to interact if and only if they
share opinions on (strictly) more than $\vartheta$ out of $F$
features. To soften this constraint we follow \cite{Flache1} and
introduce a source of noise, and allow agents who agree on $\vartheta$
or fewer features to interact with probability $\delta$.  We refer to
this type of stochasticity as `interaction noise' in the following,
$\delta$ measures its strength. $\gamma$ in the above update rules
instead denotes the strength of what we will call `external
noise'. After each time step, with probability $\gamma$ a randomly
chosen component of a randomly chosen spin is set to a random value
$\{1,\dots,q\}$. This type of noise has first been studied in
\cite{Klemmnoise}.

\section{Master Equation in the Mean Field approximation}

In this section we will consider a mean field approximation of the
model.  In the mean field model it is possible and convenient to
consider the dynamics in terms of bonds, i.e. of pairs of neighbouring
agents, rather than in terms of spins $\{\vec{\sigma}_i\}$.  Following
the strategy of \cite{Castellano, Vilone, Redner}, let $P_{m}(t)$ be
the probability that, at a given time $t$, a bond is of type $m$,
i.e. that the two agents at the ends of the bond have the same opinion on
exactly $m$ features. We will occasionally refer to bonds of type $F$
as `fully saturated' in the following. If we let $\rho$ be the
probability that at the starting point of the dynamics two spins share
a given feature, we have initially
\be
P_{m}(t=0)=\binom{F}{m}\rho^{m}(1-\rho)^{F-m}
\ee
For $\sigma_{if}$ drawn independently and with equal 
probabilities from $\{1,\dots,q\}$ one has $\rho=1/q$. 

We further define $\lambda$ to be the probability that two independent
spin components are equal but different from a given third. 
$\lambda$ is in principle
a time-dependent quantity as the system evolves according the Axelrod
dynamics. We here neglect this time-dependence and assume that
$\lambda$ is well approximated by its initial value
$\lambda=(q-1)^{-1}$ throughout the dynamics. This was seen not to
have any significant effects on results in \cite{Castellano,Vilone}. 
Now denote by $W^{(k)}_{n,m}$ the transition probability that
a bond of type $n$ becomes of type $m$ due to the updating of a
neighboring bond of type $k$.  The only non-zero elements are \cite{Castellano, Vilone, Redner}
\[
\begin{array}{lll}
 W^{(k)}_{n,n-1} & = & n/F,  \\
 W^{(k)}_{n,n} & = & (1-\lambda)\left(1-n/F\right), \\
 W^{(k)}_{n,n+1} & = & \lambda\left(1-n/F\right) \ , 
\end{array}
\]
independently of $k$. We will therefore suppress the 
superscript $k$ in the following.
 
Let us further define $I_k$ to be the probability with which 
two agents who share opinions on $k$ features interact if 
selected for potential update. Then one has 
\begin{equation*}
I_k=\left\{\begin{array}{cc} \delta &~~~~  k\leq\vartheta \\ k/F &~~~~ \vartheta
+1\leq k<F \\ 0 &~~~~ k=F.\end{array}\right.
\end{equation*}
The master equation can then be written in the form
\BE
\frac{g}{2}\frac{dP_{m}(t)}{dt}&=&\sum_{k=0}^{F-1} 
\left[ \delta_{m,k+1}-\delta_{m,k}\right] I_k P_k \nonumber \\
&& + (g-1)\left(\sum_{k=0}^{F-1}I_k P_{k}\right)
\sum_{n=0}^F (P_n W_{n,m}-P_m W_{m,n}) \nonumber \\
&&+ \gamma g \left[ (1-\delta_{m,f})P_{m+1}
\frac{m+1}{F}\left(1-\frac{1}{q}\right)- 
(1-\delta_{m,f})P_{m}\frac{m}{F}\frac{1}{q}\right. \nonumber \\
&&\left.
+(1-\delta_{m,0})P_{m-1}\left(1-\frac{m-1}{F}\right)\frac{1}{q}
- (1-\delta_{m,0})P_{m}\frac{m}{F}\left(1-\frac{1}{q}\right)\right].\label{eq:master}
\EE
This equation is an approximation in the mean-field sense, and the
thermodynamic limit is implied. The Master equation can be expected to
describe the system at most at large system sizes, and will therefore
not be able to capture features characteristic of finite systems. The
geometry of the square lattice is mimicked, in the mean-field spirit,
by the pre-factors $g-1$ and $g$ in the different terms of the Master
equation. $g$ here denotes the co-ordination number of each spin so
that these coefficients reflect the number of spins with whom a given
spin can interact (albeit these are not nearest neighbors any
longer). On a square lattice in two dimensions one has $g=4$. The
pre-factor $g/2$ in front of the time derivative in
Eq. (\ref{eq:master}) takes into account the fact that the system
contains $g/2$ bonds per lattice site. One would expect the Master
equation to be accurate in the case of degree-regular graphs (of
connectivity $g$), as discussed for example in \cite{Redner}. Still,
as demonstrated \cite{Castellano} and as we will see below in the
context of external and interaction noise, an approach based on
numerical integration of the Master equation is able to reproduce some
features of the two-dimensional model at least qualitatively.


\section{Axelrod dynamics with nominal features}
\subsection{Baseline model}
\begin{figure}
\vspace{1em}\includegraphics[width=0.35\textwidth]{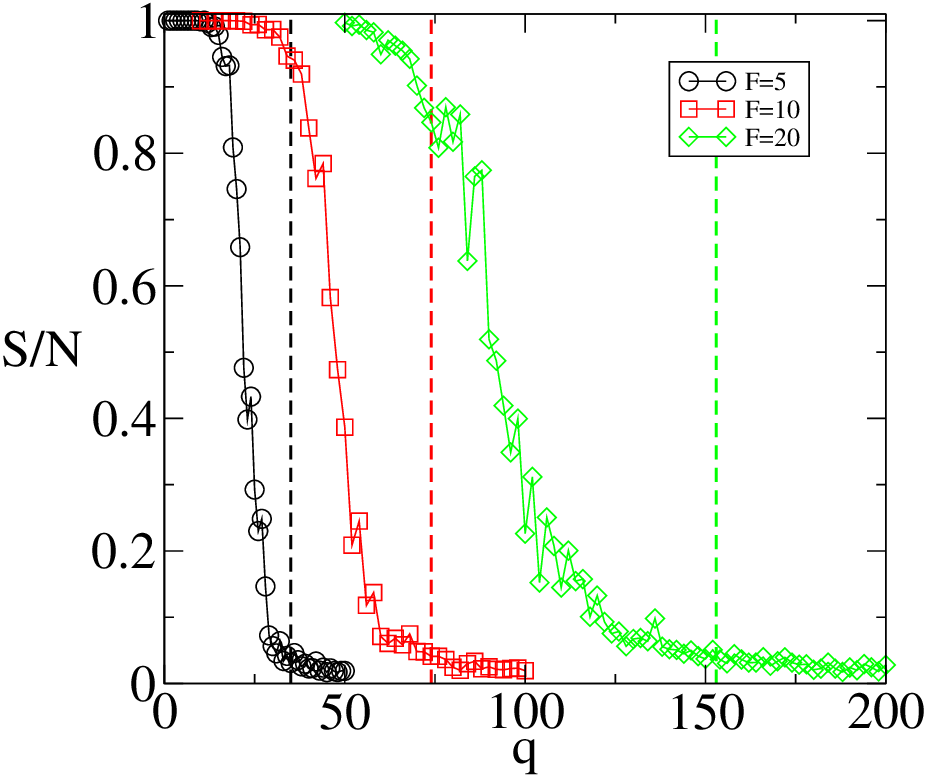}~~~~~\includegraphics[width=0.35\textwidth]{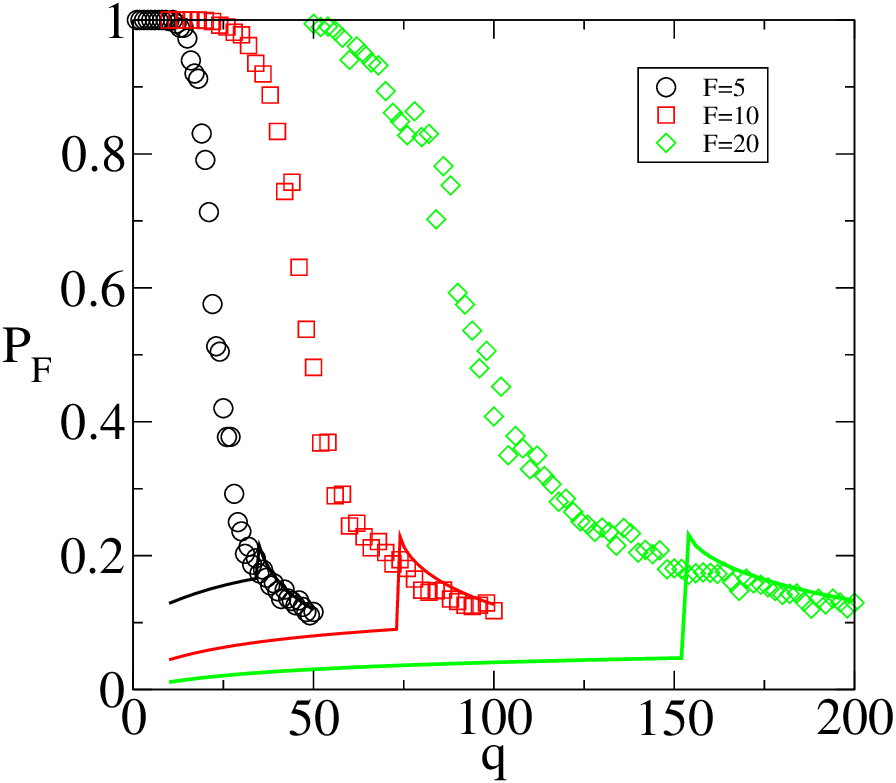}
\caption{\label{fig:baseline} (Colour on-line) Relative 
size $S/N$ of the largest cultural region in the standard 
Axelrod model with $F$ nominal features, 
initial conditions drawn from $\sigma_{if}\in\{1,\dots,q\}$ with flat distribution. 
Symbols show data from simulations run until convergence, 
system size $N=400$, averaged over $10$ samples, $F=5,10,20$. 
The vertical dashed lines mark the location of the phase 
transition as predicted by a numerical integration of the Master equation. 
The right panel shows $P_F$ from simulations (symbols) 
compared to results predicted by the Master equation (solid lines).}
\end{figure}

For completeness we re-iterate the behaviour of the baseline Axelrod
model ($\theta=\gamma=\delta=0$) in Fig. \ref{fig:baseline}. For any
given number $F>2$ of features a discontinuous transition between an
ordered state at $q<q_c(F)$ and a disordered phase at larger values of
$q$ is observed. At $q<q_c$ the coarsening dynamics of the model
persists until a fully ordered state is reached. For any feature
$f=1,\dots,F$ all agents then agree on one opinion,
i.e. $\sigma_{if}=\sigma_{jf}$ for all $i,j$. In finite systems such a
state is reached after a finite time. Fig. \ref{fig:baseline} depicts
the relative size $S/N$ of the largest culturally homogeneous region
of spins as a function of $q$. A homogeneous region $R$ is here
defined as a subset of the $L\times L$ agents, so that within $R$ all
agents agree on all features
\cite{comment}. As seen in the figure, one finds only one region at
convergence for $q<q_c$, and has $S/N=1$. At values of $q$ larger than
$q_c(F)$ dynamic arrest occurs before the system can reach a fully
ordered phase. After the arrest no further ordering is possible due to
the kinetic constraints imposed on the otherwise attractive
spin-dynamics. The system remains in a disordered state, marked by a
large number of small cultural regions and a vanishing number of
active bonds. $S/N$ remains small at convergence in this regime. As
seen in Fig. \ref{fig:baseline} the fraction $P_F$ of fully saturated
bonds behaves similarly to $S/N$ at convergence, and can be well
captured by the Master equation in the disordered regime. The ordering
at low values of $q$ can not be obtained from an approach based on the
Master equation \cite{Castellano}.

\subsection{Effects of interaction threshold}
\begin{figure}
\vspace{3em}\includegraphics[width=0.35\textwidth]{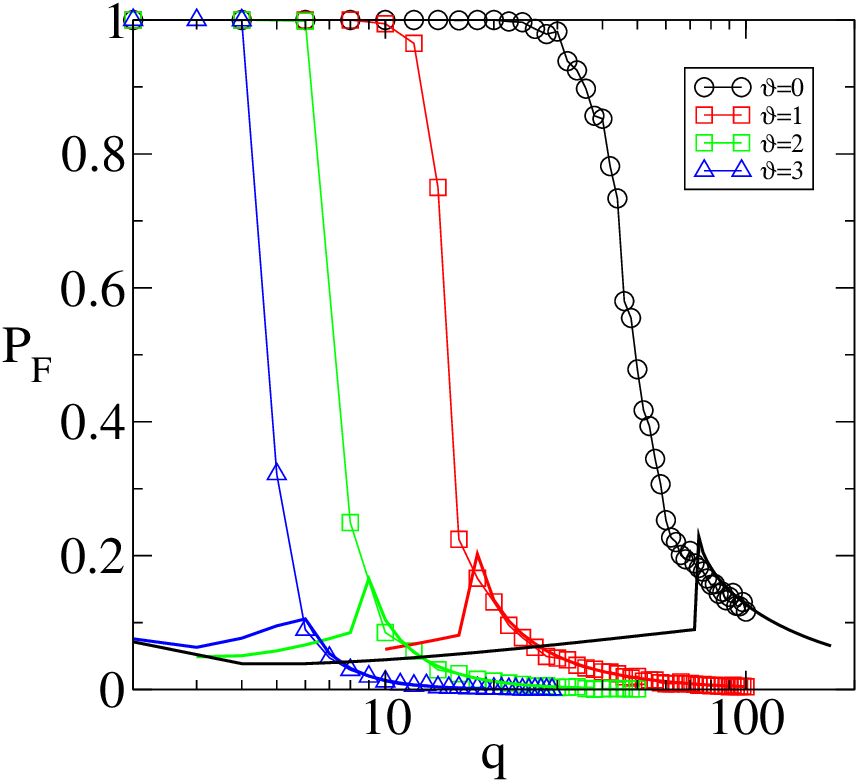}
\caption{\label{fig:thresh} (Colour on-line) Fraction of fully
  saturated bonds $P_F$ in the Axelrod model with $F=10$ nominal
  features and with confidence threshold $\vartheta=3,2,1,0$ (from
  left to right), initial conditions drawn from
  $\sigma_{if}\in\{1,\dots,q\}$ with flat distribution. Connected
  markers are from simulations with $N=400$ agents, run until
  convergence, averaged over $10$ samples.  Thick solid lines show
  theoretical predictions by the Master equation.}
\end{figure}

We now turn to the noise-free model with general interaction threshold
$\vartheta$. As shown in Fig. \ref{fig:thresh} the qualitative
behaviour of the model is not affected by the introduction of a
confidence threshold. As before an ordered phase is found at low
values of $q$, and a disordered one at $q$ larger than some critical
value $q_c(\vartheta,F)$.  One finds that an increased confidence
threshold suppresses interaction and hence reduces the range of $q$ in
which order can be reached. $q_c$ is a decreasing function of
$\vartheta$ at fixed $F$. Fig. \ref{fig:thresh} also demonstrates that
the Master equation given above describes the qualitative behaviour of
the system and dependence on the interaction threshold
appropriately. In the disordered phase even a reasonable quantitative
agreement between numerical measurements of the fraction of fully
saturated bonds and the corresponding theoretical predictions can be
observed.  We attribute remaining discrepancies to the mean field
approximation, inaccuracies in capturing the $2$-dimensional geometry
and to finite-size effects. 

\begin{figure}[t]
\vspace{2em}
\includegraphics[width=0.35\textwidth]{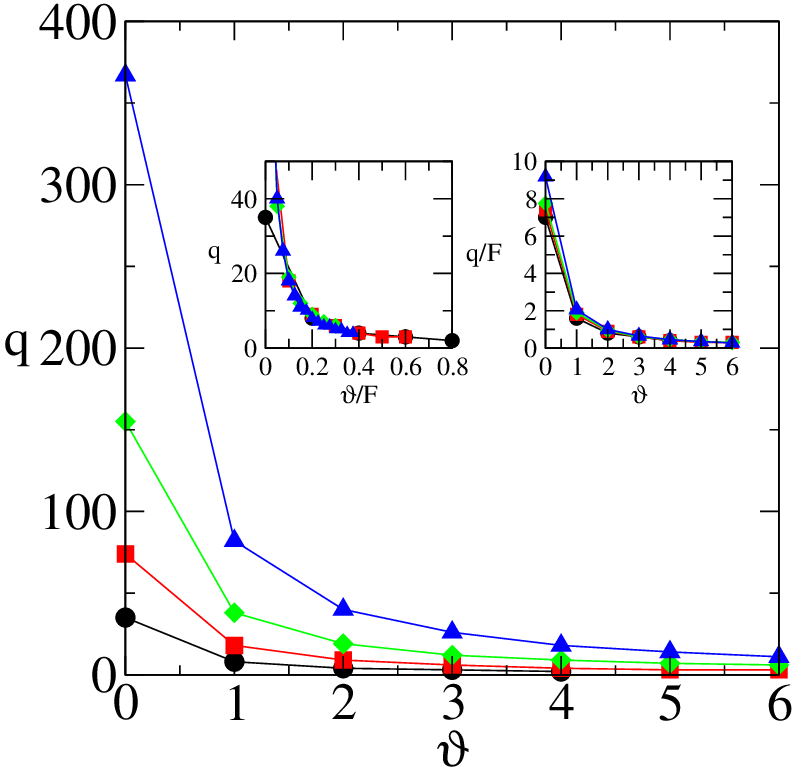}
\caption{\label{fig:pg_thresh}  (Colour on-line) Phase diagram of the model with general
  confidence threshold in the mean-field approximation. Locations of
  the transitions are obtained from the Master equation. The curves
  show $F=40,20,10,5$ from top to bottom. The system is in the
  disordered phase above the respective lines. Ordered states can be
  expected below. The insets show a rescaling in terms of the relative
  threshold $\vartheta/F$ (left) and of the relative number of opinion
  states per feature $q/F$ (right).}
\end{figure}

Fig. \ref{fig:pg_thresh} depicts the phase diagram of the model in the
$(\vartheta,q)$ plane for different values of $F$, as obtained from
the Master equation \cite{pgnum}. The disordered phase is found at
large values of $\vartheta$ and $q$ respectively, order is reached at
low $q$ and/or $\vartheta$. It may here be interesting to ask whether
the relevant variable is the absolute interaction threshold
$\vartheta\in\{0,\dots,F\}$, or the relative one $\vartheta/F$.  In
\cite{Flache1,Flache2} results are for example reported in terms of
relative thresholds. The left inset of Fig. \ref{fig:pg_thresh}
confirms that the phase boundaries for different values of $F$ as
shown in the main panel do indeed show a reasonable collapse if
plotted as a function of $\vartheta/F$.  At small values of
$\vartheta$ systematic deviations are however observed.  A different
rescaling was suggested in \cite{Maxi2}, where results for the
one-dimensional Axelrod model where shown to depend mostly on $q/F$.
As demonstrated in the right inset of Fig. \ref{fig:pg_thresh} equally
good collapse is observed in the $(\vartheta/F,q)$ plane, so that we
can here not reach a definitive conclusion as to whether there are any
independent scaling parameters, and if so which ones they are
\cite{comment2}.

\begin{figure}[t]
\vspace{2em}
\includegraphics[width=0.35\textwidth]{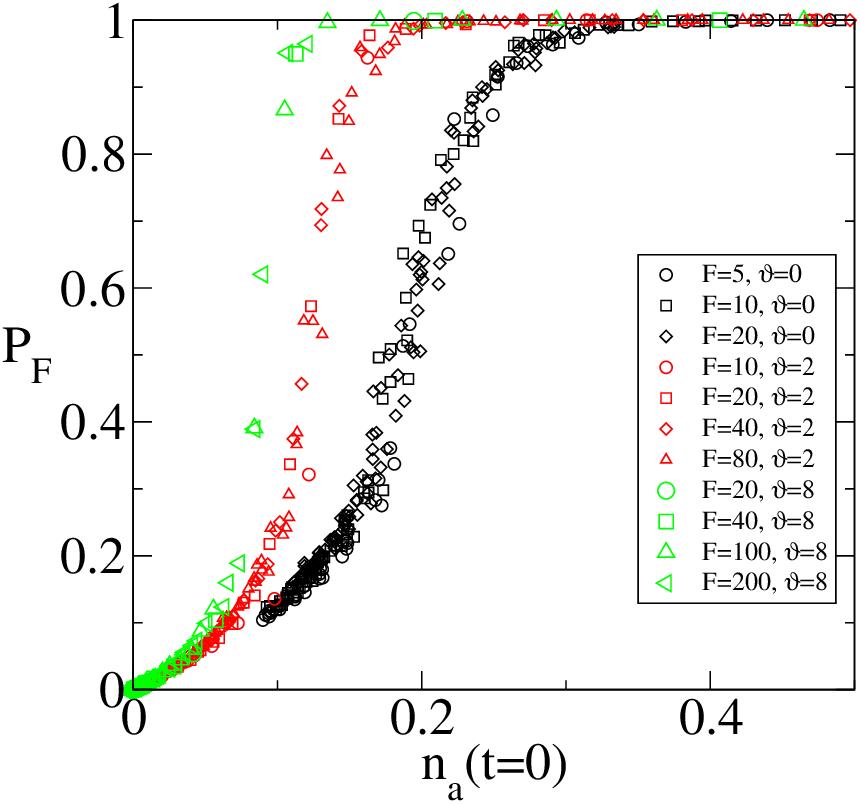}
\caption{\label{fig:threshcollapse} (Colour on-line) $P_F$ versus
  fraction of active links $n_a(0)$ in initial configuration for
  nominal Axelrod dynamics at different $\vartheta$ and $F$. Data are
  from simulations (system size $N=20\times 20$, run until
  convergence, averages over $10$ samples are taken).}
\end{figure}

Some indications regarding the relevance of {\em absolute} as opposed
to {\em relative} thresholds can be found in
Fig. \ref{fig:threshcollapse}, where we show the density of fully
saturated bonds $P_F$ as a function of the density of initially
active bonds $n_a(0)=\sum_{k=\vartheta+1}^{F-1} P_k(t=0)$.
Simulations are here performed by fixing $\vartheta$ and $F$ and by
subsequently varying $q$. $n_a(0)$ then decreases with increasing
$q$. The data shown in the figure suggest a potential collapse on three different
curves, one for each of the tested values $\vartheta=0,2,8$. While these findings
might indicate some potential universality as $F$ and $q$ are varied
at fixed $\vartheta$, reaching a final conclusion as to whether
absolute or relative thresholds are the relevant ones still remains an
open question.

\subsection{Effects of external noise}
We now turn to a discussion of the effects of noise on nominal Axelrod
dynamics. We will first focus on external noise as introduced
above. This type of stochasticity describes random fluctuations which are
triggered by an external event, and which result in individual spin
components being flipped randomly at a given rate. External noise
was introduced in the context of the Axelrod model in
\cite{Klemmnoise}, and the resulting random mutations describe what
Axelrod refers to as `cultural drift'
\cite{Axelrod} in the population of agents. Klemm et al.
\cite{Klemmnoise} have studied the effects of external noise as a
function of the noise rate and system size in fully equilibrated
systems. We here extend this analysis, and consider different cases
according to whether the thermodynamic limit or long-time limit is
taken first. Results from an integration of the Master equation are
discussed in order to provide a semi-analytical description of the
system in the limit of infinite size.

\subsubsection{Finite system, equilibrated dynamics}
In {\em finite} systems a continuous transition between an ordered
state at low noise rates $\gamma$ and a disordered state at large
$\gamma$ has been identified in \cite{Klemmnoise}. This transition
relates to a characteristic relaxation time $T={\cal O}(N\log N)$ in
finite systems. When the noise rate is sufficiently large (larger than
$T^{-1}$) stochastic perturbations build up in time, and lead to a
disordered state. For $\gamma T << 1$ the system drifts from one
ordered state to another in time, time-averaging effectively yields
global order. Since at $q>q_c$ disorder is observed in the absence of
noise ($\gamma=0$), the behaviour of the model in the limit $\gamma\to
0^+$ is discontinuous at $q>q_c$. As discussed in the next point the
finiteness of the system is crucial here, so that the described
behaviour cannot be captured by the Master equation.

\begin{figure}[t]
\vspace{6em}
\includegraphics[width=0.6\textwidth]{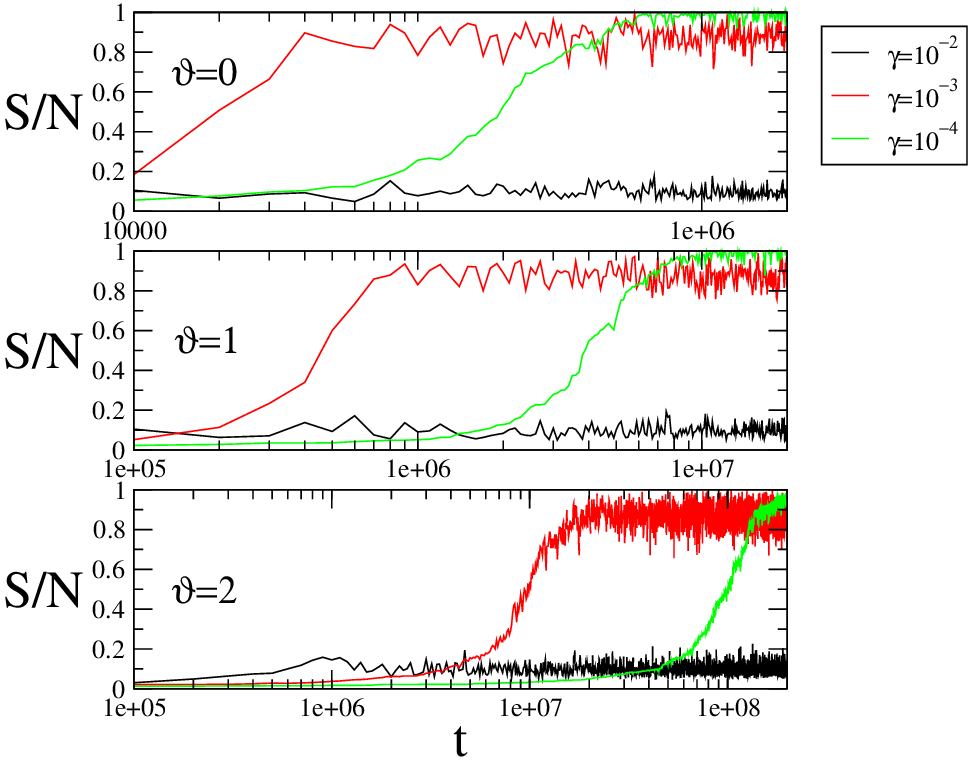}
\caption{\label{fig:noise_thresh_time} (Colour on-line) $S/N$ versus time for $F=10,
  q=100$. $\vartheta=0,1,2$ in top, middle and lower panel. Curves
  shown for $\gamma=10^{-2}, 10^{-3}, 10^{-4}$ (bottom to top at large times). From simulations,
  $N=10\times 10$, averages over $10$ runs. The $x$-axes show time in
  multiples of $N$ microscopic updates, i.e. after $t$ every agent has
  on average been selected $t$-times for potential update.}
\end{figure}

The effects of external noise on {\em finite} Axelrod systems with
general threshold is depicted in Fig. \ref{fig:noise_thresh_time}. The
behaviour of the model with confidence threshold is here found to be very similar
to the one identified in \cite{Klemmnoise} for conventional Axelrod
dynamics. We here focus on $F=10$, $q=100$ as an example, but similar
behaviour can be expected for other model parameters in the disordered
phase of the noise-free model. For small values of $\gamma$ the system
orders after an initial transient. At large magnitude of the applied
noise, no ordering is found, consistently with the results of
\cite{Klemmnoise}. This general qualitative picture appears to be
independent of the applied threshold.  The duration of the
equilibration period in cases where the system orders, however, shows
a significant dependence on the noise strength and on the chosen
threshold. Generally, the time required to reach equilibration
increases as $\gamma$ is lowered or as $\vartheta$ is increased, see
Fig. \ref{fig:noise_thresh_time}. The value of $S/N$ at equilibrium is
a decreasing function of $\gamma$. In the examples of
Fig. \ref{fig:noise_thresh_time} equilibration to a value of
$S/N\approx 0.1$ occurs fast at $\gamma=10^{-2}$. At lower noise
rates, $S/N$ reaches values in the range of $0.8$ to $1$, but only
after a substantial equilibration period, which increases as $\gamma$
is lowered. Only models with low or moderate threshold and/or
sufficiently large noise strength can hence be equilibrated in
reasonable computing time. While analogy suggests that an ordered
phase sets also at higher thresholds $\vartheta$ and small enough
noise strengths if the dynamics is run long enough, we have not been
able to confirm this explicitly due to computational
limitations. Approaches based on continuous-time Monte Carlo methods
might here potentially be more appropriate than direct simulation of
the Axelrod dynamics \cite{newman}.

\subsubsection{Thermodynamic limit for equilibrated systems}
Due to the divergence of the relaxation time $T$ with $N$, the
transition just discussed disappears in infinite systems. 
If, at $q>q_c$, the system size is
taken to infinity {\em after} equilibration, i.e. if equilibrated systems are considered at increasing $N$ while keeping all other parameters fixed, the population always ends up in
a disordered state for all $\gamma\geq 0$, as demonstrated in
\cite{Klemmnoise}. The discontinuity at $\gamma\to 0^+$ is removed.  
Equilibrated systems of infinite size cannot be captured by the
Master equation, as the latter implies the thermodynamic 
limit to have been taken first.

\subsubsection{Thermodynamic limit at finite $t$}
\begin{figure}[t]
\vspace{2em}
\includegraphics[width=0.4\textwidth]{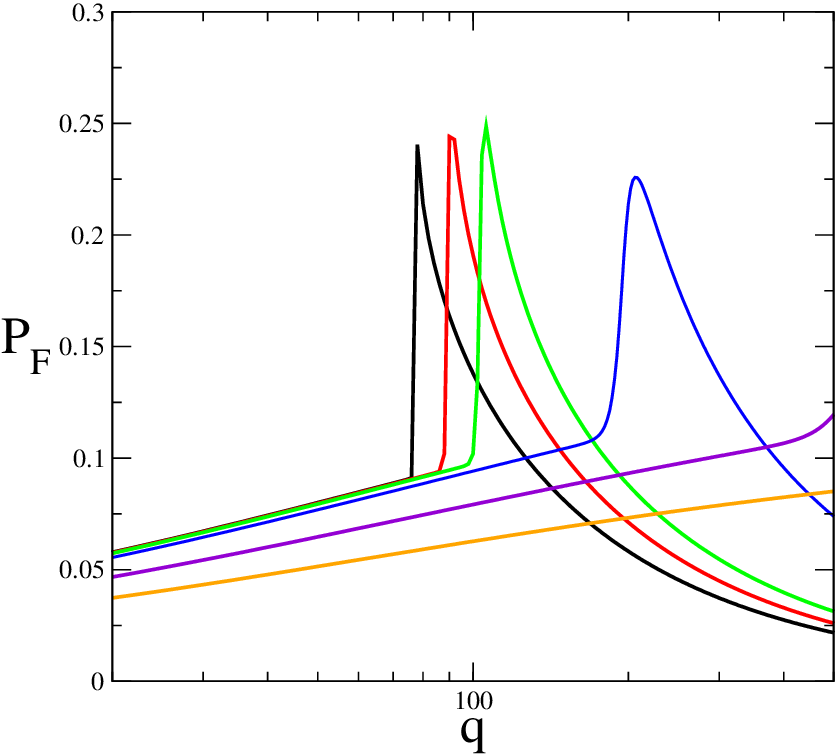}
\caption{\label{fig:poffq}  (Colour on-line) $P(F)$ versus $q$ in a nominal Axelrod
  system, $F=10$. Results are from numerical integration of the Master
  equation up to $t=5000$. Curves show
  $\gamma=0.0001,0.0005,0.001,0.005,0.025,0.05$ from left to right at the
  maximum. Curves for $\gamma=0.025$ and $\gamma=0.05$ still display a
  maximum, but to the right of the plotted range of $q$.}
\end{figure}

Taking the thermodynamic limit $N\to\infty$ at a fixed number of
macroscopic time steps $t$ results in non-trivial behaviour. As
displayed in Fig. \ref{fig:poffgamma} one finds a disordered state
with $P_F\approx 0$ at sufficiently large $\gamma$. Partial ordering
sets in as $\gamma$ is lowered, with non-monotonous behaviour and a
peak of $P_F$ at intermediate noise-strengths. As $\gamma$ is reduced
further, $P_F$ takes small but non-zero values, mostly independent of
the noise strength, provided the latter is small enough. Integration of
the Master equation confirms this behaviour qualitatively. The
non-monotonic behaviour of $P_F$ as a function of $\gamma$ at fixed
time-scale $t$ is here potentially related to the non-monotonic
temporal behaviour of the Axelrod system as observed for example in
\cite{Castellano,Redner}. Under suitable conditions order parameters
such as $P_F(t)$ or $S(t)/N$ might become non-monotonic functions of
time at fixed values of $q,F,\gamma$. These non-monotonicities are
then reflected as peaks in $P_F$ when other cuts through parameter
space are considered, as in Fig. \ref{fig:poffgamma} where $\{t,F,q\}$
are fixed and $\gamma$ is varied, or in Fig. \ref{fig:poffq} where $q$
is varied at fixed $\{F,t,\gamma\}$.

\subsubsection{Large time limit {\em after} taking thermodynamic limit}
\begin{figure}[t]
\vspace{2em}
\includegraphics[width=0.4\textwidth]{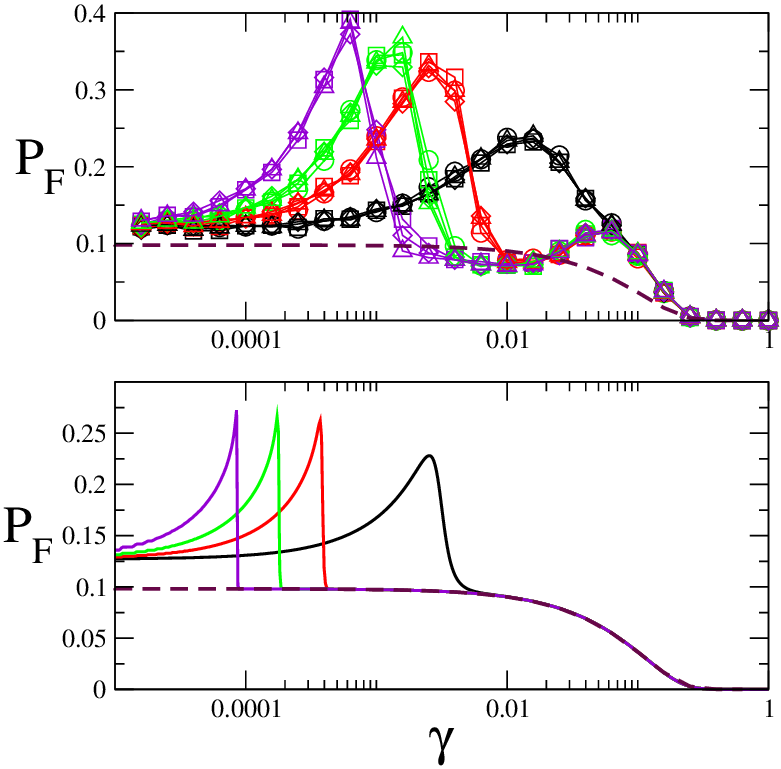}
\caption{\label{fig:poffgamma}  (Colour on-line) $P_F$ versus $\gamma$ in a nominal
  Axelrod system, $F=10, q=100$. Upper panel: results from
  simulations, run for $t$ macroscopic time-steps, where
  $t=20000,10000,5000,1000$ from left to right. Circles are for
  $N=20\times 20$, squares for $N=30\times 30$, diamonds $N=40\times
  40$, triangle $N=50\times 50$. Data are averages over $10$ samples.
  Lower panel: $P_F$ from numerical integration of the Master
  equation, $t=20000,10000,5000,1000$ as above. An Euler-forward
  scheme with time-step $dt=0.1$ is here used. Dashed line in both
  panels marks result from Master equation after even longer times
  ($t=2\cdot10^5$).}
\end{figure}

As the time-scale on which the systems is studied is increased, the
peak in $P_F$ appears to move further to the left in
Fig. \ref{fig:poffgamma}, i.e. to smaller values of the noise strength
$\gamma$. Hence for any fixed $\gamma_0$ there is a time-scale
$t(\gamma_0)$ so that $P(F)$ is monotonically decreasing as a function
of $\gamma>\gamma_0$ on this time-scale. The system is hence
disordered at large $\gamma$, and partially ordered at low
$\gamma$. This behaviour can successfully be described by the Master
equation.

\subsection{Effects of interaction noise}
We next turn to a discussion of the effects of interaction noise, as
parametrised by its amplitude $\delta$. This type of stochasticity
facilitates interaction, as it removes kinetic constraints and allows
agents sharing $\vartheta$ or less opinions to interact
(at rate $\delta$), while in standard Axelrod dynamics they would not
be able to align spin components. While the external noise of
amplitude $\gamma$ has an ambiguous role of inducing order at low
amplitudes and of driving the system to disorder at large $\gamma$,
interaction noise can generally be expected to favour order. It his hence
interesting to study the system in presence of both types of
randomness, and to identify re-inforced ordering behaviour or (at
small $\gamma$) potential competition between the ordering and
disordering stochasticity (at large $\gamma$)

\begin{figure}
  \vspace{3em}
  \includegraphics[width=0.33\textwidth]{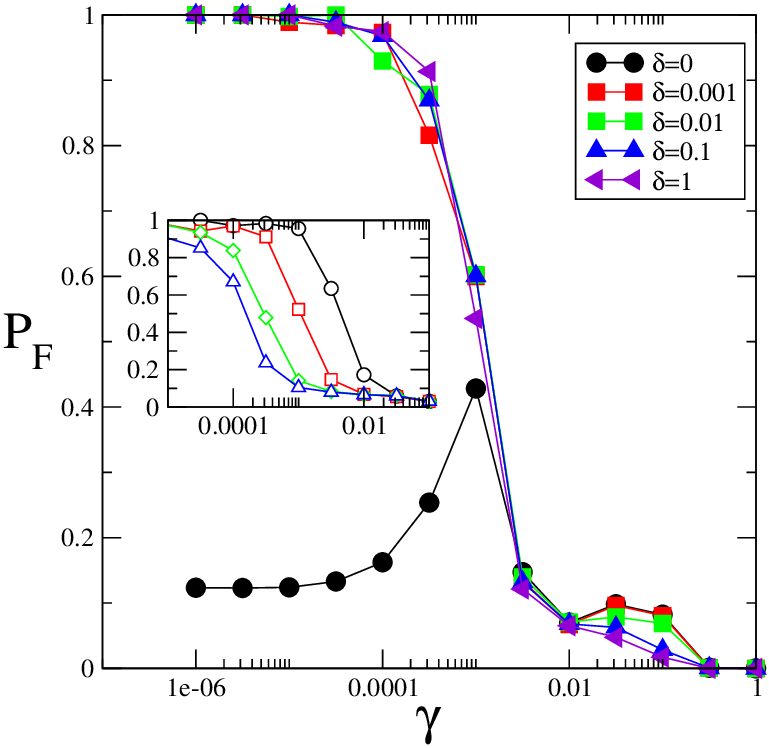}~~~~~\includegraphics[width=0.35\textwidth]{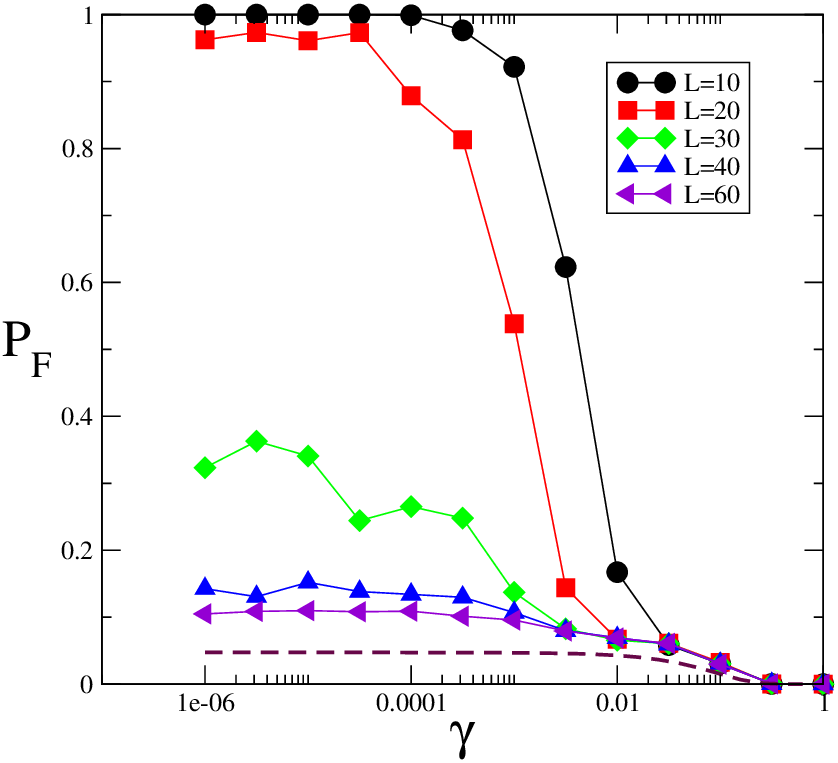}
  \caption{\label{fig:gammaprime} (Colour on-line) Effects of
    interaction noise. {\bf Left:} $P_F$ versus $\gamma$ for Axelrod
    model ($F=10,q=100,\vartheta=0$), $N=20\times 20$, run for $20000$
    steps (averages over $10-20$ samples) The inset shows results for
    systems of size $N=10\times 10,20\times20, 40\times 40, 60\times
    60$ at $\delta=0.1$ from right to left. {\bf Right:} $P_F$ for
    different system sizes at fixed $\delta=0.1$ at $t=5000$. Dashed
    line in right panel is from numerical integration of the Master
    equation.}
\end{figure}

Results for a nominal Axelrod system with both types of noise are
reported in Fig. \ref{fig:gammaprime}. The data indicate that the
effects of interaction noise are mostly to facilitate order for large
ranges of fixed external noise $\gamma$. More specifically, the
effects of interaction noise is to reduce the time-scale on which
finite systems order in the presence of external noise. The left panel
of Fig. \ref{fig:gammaprime} shows the concentration $P_F$ of bonds
with full overlap in an Axelrod system run for a time which is not
long enough for the system without interaction noise (circles,
$\delta=0$) to develop order at the studied magnitudes $\gamma$. Order
at $\gamma\lesssim 0.001$ would develop only if the system were run
for longer times, and had fully equilibrated. The curves for
non-vanishing amplitude $\delta$ demonstrate the effect of interaction
noise, the system now orders at small noise strengths $\gamma$ . While
the qualitative behaviour of the {\em equilibrated} system is not
altered, the facilitation of the kinetic constraints drastically
reduces equilibration times, and the system orders at sufficiently low
$\gamma$ even after moderate running times. Interestingly the noise
strength $\gamma_0$ separating the ordered from the disordered regime
of the fully equilibrated system appears not to be affected much by
the interaction noise. The curves displayed in the left panel of
Fig. \ref{fig:gammaprime} are indeed mostly independent of $\delta$,
as long as $\delta>0$. It might potentially be interesting to study
even lower $\delta$, although probably unrealistic from the
sociological point of view
\cite{comment3}.

We conclude that the effect of interaction noise is to reduce
relaxation times, but that it does not alter the phase behaviour of
the model, with an ordered phase at low $\gamma$, and a disordered one
at large $\gamma$. As in the absence of interaction noise ($\delta=0$)
this transition is present only in finite systems, seen in the inset
of the left panel of Fig. \ref{fig:gammaprime} and in the right
panel. As the system size is increased at equilibrium the
order-disorder crossover moves to smaller values of the noise strength
$\gamma$, and can be expected to be absent in the thermodynamic limit,
where only the disordered region prevails. Indeed rescaling of the
data in the inset of the left panel demonstrates that $\gamma N\ln N$
is the relevant scaling variable, similar to the observations of
\cite{Klemmnoise}. At finite running times $t$ the order at low
$\gamma$ is gradually reduced with increased system size and in the
thermodynamic limit the system is qualitatively well described by the
Master equation.

\begin{figure}
  \vspace{3em}\includegraphics[width=0.35\textwidth]{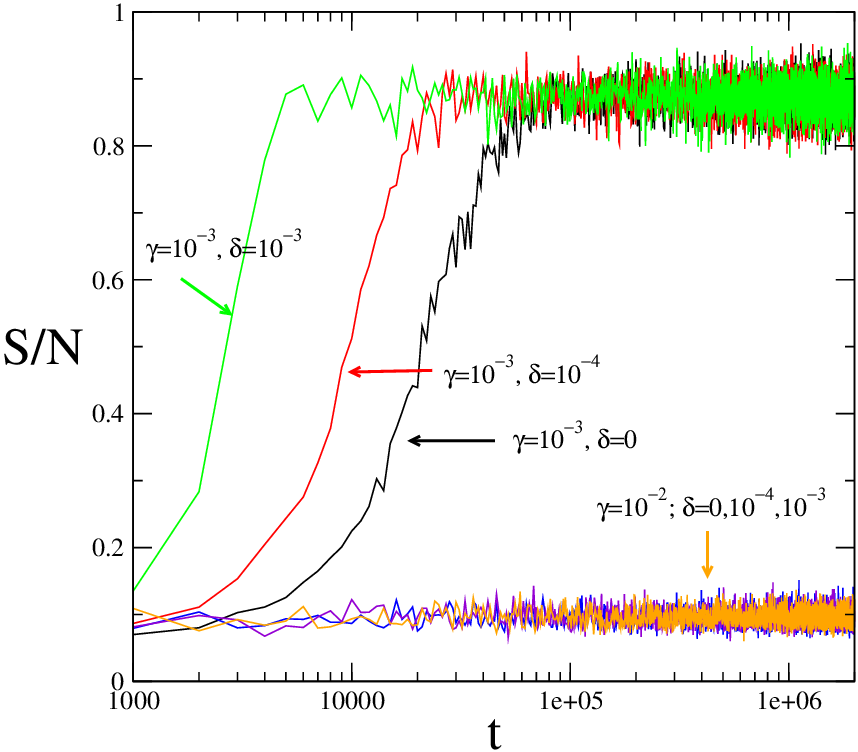}~~~~\includegraphics[width=0.35\textwidth]{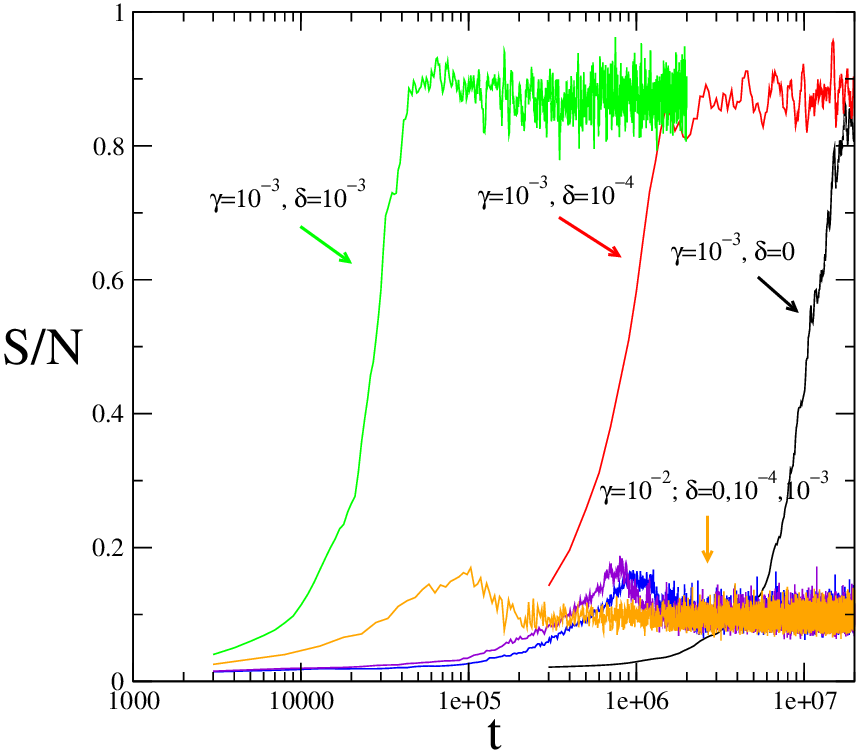}
  \caption{\label{fig:timescan_th02} (Colour on-line) Effects of
    interaction noise. $S/N$ versus time $t$ at different levels of
    external and interaction noise. Data is from simulations of
    systems with $N=10\times 10$ agents. The left part shows
    $\vartheta=0$ (averages over $50$ samples), the right one
    $\vartheta=2$ ($10$ runs, data smoothened by running averages).
    $F=10,q=100$ in both panels. At large $\gamma$ no order is
    reached, independently of $\delta$. For small enough $\gamma$ the
    system orders, the time-scale on which this occurs is largely
    reduced by the introduction of interaction noise, and increases as
    the interaction threshold is raised (note the different scaling on
    the time axes on the left and right).}
\end{figure}

Fig. \ref{fig:timescan_th02} finally confirms that this behaviour is
not limited to the standard Axelrod dynamics with vanishing
interaction threshold ($\vartheta=0$). Interaction noise reduces the
time-scale on which the system orders at small $\gamma$ also in the
model with moderate non-zero thresholds, and that the system at $\vartheta>0$
behaves very much like the one at $\vartheta=0$.

\section{Euclidean Axelrod dynamics}

We now turn to a modification of the Axelrod dynamics in which a
metric allowing for gradual notions of agreement between agents is
introduced. As before, opinions take discrete values
$\sigma_{if}\in\{1,\dots,q\}$. However, the `distance' between two
spins $\vec{\sigma}_i$ and $\vec{\sigma}_j$ is no longer measured in
terms of the number of features on which the corresponding agents
agree, but based on the following Euclidean distance between the
spin-vectors $\vec{\sigma}_i$ and $\vec{\sigma}_j$:
\be
d=\sqrt{\frac{1}{F(q-1)^2} \sum_{f=1}^F
(\sigma_{if}-\sigma_{jf})^2}.
\ee
Thus the distance between two agents ranges between $0$ and $1$. It takes the maximal value $d=1$ if and only if the opinions of the two agents are diametrically opposed, i.e of for any feature $f$ one has $\sigma_{if}=1, \sigma_{jf}=q$ or $\sigma_{if}=q, \sigma_{jf}=1$. Distances different from zero or one thus indicate partial agreement between the two agents.

In the following we will take the potential of two neighbouring agents to interact to be given by the following logit-rule \cite{logit}:
\be \label{eq:pofd}
p(d)=\frac{1}{1+e^{\beta (d-d_0)}}.
\ee
$\beta$ is a control parameter allowing for the introduction of interaction noise. The case $\beta=\infty$ here corresponds to the noise-free (zero temperature) case. If $\beta=\infty$, agents with distance $d>d_0$ are unable to interact, $p(d>d_0)=0$, whereas interaction always occurs for pairs of agents with distance $d<d_0$. $d_0$ is thus a threshold parameter, with large $d_0$ corresponding to a regime of strong confidence of agents in other people's opinions, and small $d_0$ to cases in which interaction is rare. In order to avoid confusion let us at this point stress that the role of the threshold $d_0$ is inverse to the one of $\vartheta$ in the nominal Axelrod model: large $\vartheta$ make interaction rare, whereas large $d_0$ facilitate spin updates. 

Choosing finite values of $\beta$ turns the hard threshold into a
soft one. Interaction rates decrease smoothly with increasing
distance. Crucially, at finite $\beta$, interaction is always possible
in principle, even if $d>d_0$. For $\beta=0$ finally, interaction is
fully stochastic and independent of $d$. At any iteration, any chosen pair of
neighbouring agents interacts with probability $1/2$.

Let us summarise the resulting dynamics:

\begin{enumerate}
\item Select one spin $i\in\{1,\dots,N\}$ at random. Subsequently select one of its four nearest neighbours at random. Call this second spin $j$.
\item Compute the Euclidean distance $d(i,j)$ between $i$ and $j$.
\item If $d(i,j)=0$ both agents agree on all features. Interaction has no effect. If $d\equiv d(i,j)>0$ then with probability $p(d)$ as defined above spins $i$ and $j$ interact as in the nominal Axelrod model:  one feature $f$ is chosen at random so that $\sigma_{if}\neq\sigma_{jf}$. Then set $\sigma_{if}=\sigma_{jf}$.
\item External noise. With probability $\gamma$ perform
the following: choose one spin $i$ and one feature $f$ at random. Set
$\sigma_{if}$ to a value chosen randomly from $\{1,\dots,q\}$.
\item Resume at (i).
\end{enumerate}

\subsection{Noise-free dynamics}
\begin{figure}
\vspace{1em}\includegraphics[width=0.4\textwidth]{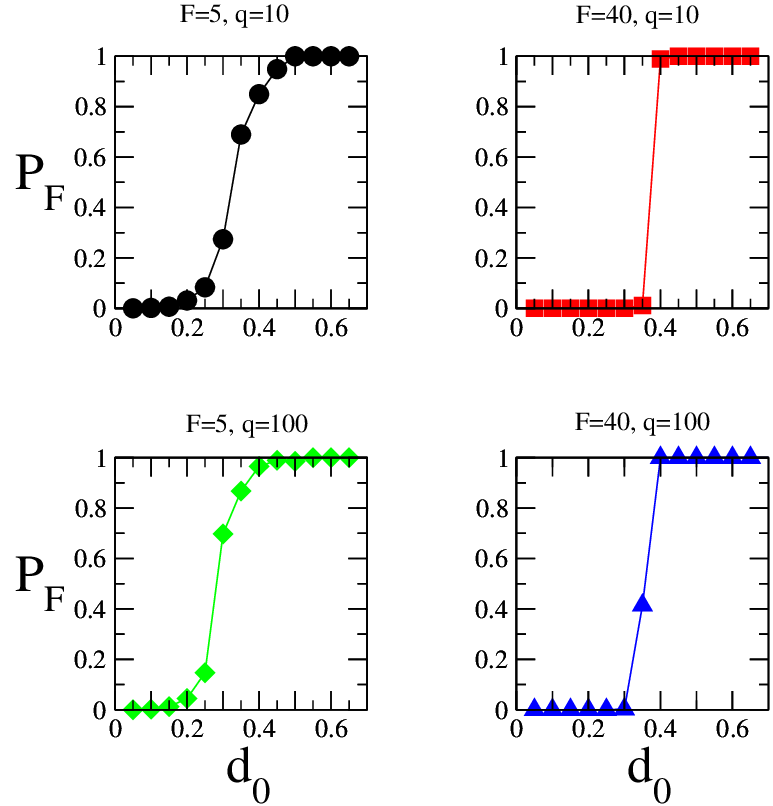}
\caption{\label{fig:poffvsdnull}  (Colour on-line) $P_F$ versus interaction threshold $d_0$ for the Euclidean model. Data is from simulations at different values of $F$ and $q$ ($N=20\times 20$, averaged over $10$ samples). Simulations are run until each bond has distance $d=0$ or $d>d_0$, i.e. until no active bonds are left in the system.}
\end{figure}

The behaviour of the noise-free system with Euclidean metric is
described in Fig. \ref{fig:poffvsdnull}. A transition between a
disordered phase at low thresholds $d_0$ and an ordered state at
larger values of $d_0$ is observed. The behaviour in these two phases
is as follows: at low $d_0$ only neighbouring agents with small
differences in opinion can interact, so that the fraction of active
bonds initially contained in the system is small. Dynamic arrest
occurs quickly, and the system remains disordered. At large enough
thresholds $d_0$ the coarsening dynamics can persist until a fully
ordered state is reached. Interestingly, as shown in
Fig. \ref{fig:poffvsdnull}, the critical value of the threshold
$d_{0c}$ does not depend much on the choice of $F$ and $q$, and takes
values $d_{0c}\approx 0.4$. The transition appears to become sharper at larger values of $F$ (see Fig.  \ref{fig:poffvsdnull}).

Plotting $P_F$ versus $q$ at fixed $F$ suggests that $q$ plays no
significant role in the Euclidean model. Only for small values of $q$
can a dependence of $P_F$ on $q$ be detected. This invariance is
intuitively to be expected as $d$ is normalised to range between $0$
and $1$ in the setup chosen here, so that $q$ is merely a measure for
how many discrete values can occur inbetween. Simulations with
continuous opinions ranging in the interval $[0,1]$ (not shown here)
reveal a behaviour very similar to the one depicted in
Fig. \ref{fig:poffvsdnull}. We have also tested models with continuous
opinions, in which both interacting agents agree on the {\em average}
opinion of a given feature in case of interaction (with the same
metric and kinetic constraints as before), and find similar behaviour
as a function of $d_0$. Similar models are discussed in
\cite{Redner2,Redner3}, mostly focussing on the case of one feature.

\subsection{Effects of external and interaction noise}

\begin{figure}
\vspace{3em}\includegraphics[width=0.35\textwidth]{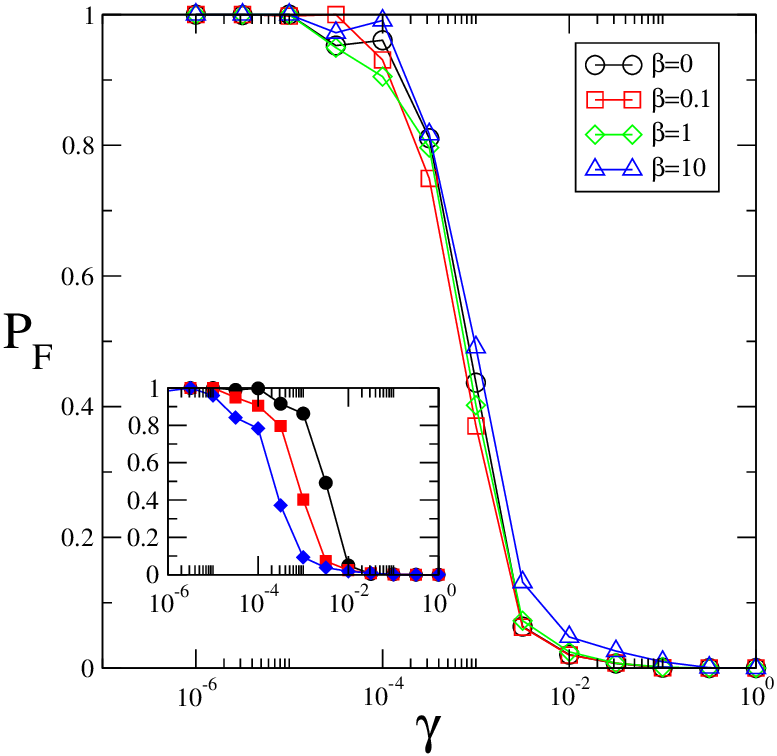}
\caption{ (Colour on-line) Effects of noise on the Euclidean model. Main panel shows $P_F$ versus $\gamma$ from simulations of a system of size $N=20\times 20$, run for $200000$ macroscopic steps, averages over $10$ samples ($F=10,q=100$). The interaction threshold is $d_0=0.3$, i.e. the system is in the disordered phase if $\beta=\infty$. Inset: $P_F$ versus $\gamma$  at fixed $\beta=1$ for system sizes $N=40^2,20^2,10^2$ from left to right (same simulation parameters as in main panel, $5$ samples only for $N=40^2$). \label{fig:euclid_gamma} }
\end{figure}

The behaviour of the Euclidean system under the influence of external
and interaction noise is shown in Fig. \ref{fig:euclid_gamma}. As seen
in the main panel, a crossover between an ordered regime at low
magnitudes $\gamma$ of the external noise and a disordered state at
higher noise-amplitudes is found, very much like in the nominal
Axelrod model. Interaction noise (finite $\beta$) appears to have only
little effect on this crossover for all values tested. Due to long
equilibration times we have not performed a full analysis of the
impact of external noise ($\gamma>0$) in the large-$N$ limit of the
model zero-temperature ($\beta=\infty$). The inset of
Fig. \ref{fig:euclid_gamma} however demonstrates that at finite
$\beta$ the range of $\gamma$ in which the system orders is reduced as
the system size is increased, similarly to what is found in the
nominal Axelrod model. The ordering behaviour at small values of the
external noise strength hence again appears to be present only in
finite systems.

\subsection{Extremism and polarisation}

One particular potential application of metric Axelrod models is to
study extremism, and the effects of a small number of highly polarised
agents, e.g. with opinions at the extreme ends of the political
spectrum. Having studied the effects of uniform confidence thresholds
on the opinion dynamics, we here extend the model to the case of {\em
heterogeneous}, i.e. agent-specific thresholds. Here any agent
$i\in\{1,\dots,N\}$ holds an individual threshold $d_{0,i}$. In an
interaction with a neighouring agent $j$ he adjusts his opinion vector
$\sigma_i$ only if $d(i,j)<d_i$, i.e. $d_0$ in Eq. (\ref{eq:pofd}) is
replaced by $d_{0,i}$. Agents with large $d_{0,i}$ are thus likely to
interact with others, and have a large tolerance against opinions in
their surroundings. Agents who are unlikely to modify their own
opinion vector are described by small interaction thresholds. Related
work on other opinion dynamics models, mostly with continuous opinions
and focusing on one feature or on nominal features, can be found in
\cite{Weisbuch1,Weisbuch1a,Weisbuch2, Fanelli}, see however also \cite{Stauffer}. Our simulations here focus on the effects of extremists in the context of a multi-variate opinion dynamics model ($F>1$) with discrete opinions and metric features.

\begin{figure}
\vspace{1em}\includegraphics[width=0.3\textwidth]{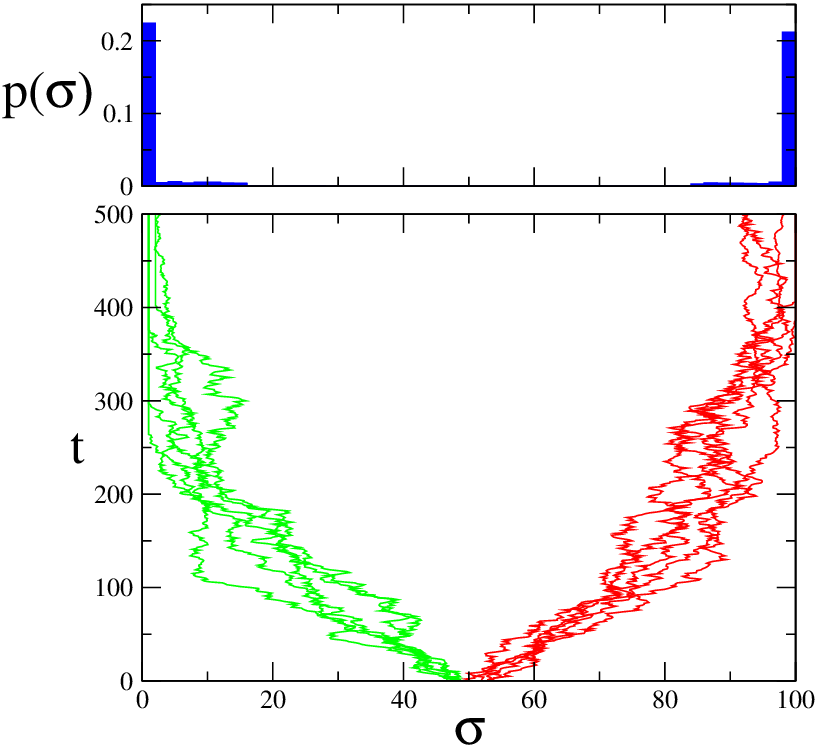}~~~~~\includegraphics[width=0.3\textwidth]{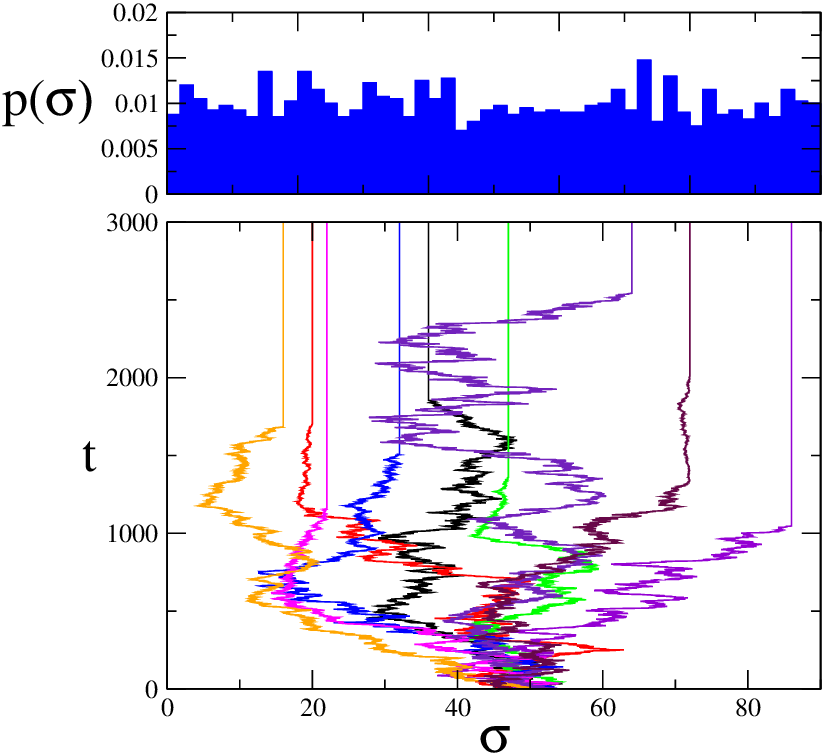}
\caption{ (Colour on-line) Metric model with (left) and without extremists (right). In the left panel a fraction of $\varepsilon=0.05$ of extremists was used, with $d_e=0.05$. $d_0=0.6$ as well as $F=10, q=100$ in both panels. Each panel shows the average opinion $N^{-1}\sum_i \sigma_{if}$ on a given feature $f$ as function of time for $10$ individual runs of a system with $N=20\times 20$ individuals, started from random initial conditions. Histograms of opinions at convergence are shown in the top part (data for $200$ samples and from all $F$ features used to generate histograms). \label{fig:extr}.}
\end{figure}

In this section we assume that the population of agents contains a
fraction $\varepsilon$ of what we will call {\em extremists}. These
are agents whose opinion vectors take extreme values $\sigma_{if}=1$
or $\sigma_{if}=q$ at the beginning of the dynamics, and who are
intolerant against other agents' opinion. In particular we assume an
interaction threshold small $d_e$ for such agents. All other agents
are taken to have a uniform threshold $d_0$ as before and are
initialised at random opinion vectors. The choices in the simulations
presented in Fig. \ref{fig:extr} are $\varepsilon=0.05$, $d_e=0.05$
and $d_0=0.6$ (the latter threshold is chosen to ensure the system is
in the `active' ordered phase where dynamics persists long enough to
prevent the system from remaining stuck in a configuration similar to
the initial condition). The figure shows the time evolution of the
average opinion of the population on a given feature, along with the
distribution of opinions at convergence. Apart from the initial
conditions no stochasticity is present in the simulations shown in
Fig. \ref{fig:extr}, i.e. we have $\beta=\infty$, $\delta=0$. As shown
in the left panel a small group of extremists can polarise the
population, provided they are inert enough against adapting their own
opinions. The figure shows $10$ runs of the Axelrod dynamics. In each
run extremists are chosen either to correspond to opinion states $1$
or $q$. Extensions to two groups of extremists at either end of the
political spectrum, or to cases in which extremists have polarised opinions only on some but not all features, are straightforward. The right panel shows the
behaviour of the system in absence of extremists. Here the average
opinion on each feature converges to a random value between $1$ and
$q$ depending on the stochastic initial conditions, and the resulting
histogram of final opinions is flat.

\section{Summary and conclusions}
We have extended Axelrod model for social influence to include varying
interaction threshold, noise and metric features. In the basic model,
which is two-dimensional, individuals are represented as
multi-component spins and interact to become culturally closer,
starting from a disordered random initial state. In our extensions
some external and interaction noises are introduced, a confidence
threshold limits the interaction, and a notion of distance between
opinion is considered. We find that the confidence threshold does not
influence the qualitative behaviour of the model, and that the typical
transition, separating an ordered and a disordered equilibrium state,
is preserved and determined basically by the initial probability of
interaction between two individuals. The threshold limits the
probability of interaction, and hence favors disorder. The
introduction of an external noise brings in finite systems a
continuous transition between an order-favoring and a
disorder-favoring role of the noise, according to whether the noise is
small or large respectively, independently of the threshold.  This
implies a discontinuity, in the region where the final state would be
disordered in the absence of external noise. Such a discontinuity is
removed in the thermodynamic limit. An interaction noise instead
always favors order by reducing relaxation times, and does not alter
the phase structure of the model. The other variant we study is the
one with a notion of distance between opinions.  We find that the
model exhibits an order-disorder transition as the distance threshold
is varied, consistently with the idea that the transition is the
result of the competition between an ordering dynamics (the relative
importance of which is determined by the distance threshold) and an
initial disorder (as measured by size of the space from which the
starting configuration is drawn at random). Moreover, the introduction
of heterogeneous confidence thresholds in the context of metric Axelrod
systems allows one to study e.g. the question of whether extremism can
prevail in such models.  We find that the presence of a small fraction
of individuals with a sufficiently rooted opinion can drive the whole
population to the extreme ends of the opinion spectrum. Further
application of heterogeneous interaction thresholds and metric features
might include extensions addressing immigration or geographic
barriers. Immigrants can for example be assumed to be more likely to
interact with other immigrants than with members of the original
population, and geographical barriers can be modelled by suppressing
interaction at certain locations in space. This would lead to different
interaction thresholds and tolerance levels, modulated either in space
or dependent on the two agents picked for potential interaction.

\begin{acknowledgements} This work was partially supported by EU
NEST No. 516446 COMPLEXMARKETS. The authors would like to thank Matteo
Marsili for helpful discussions.
\end{acknowledgements}


\begin{thebibliography}{99}
\bibitem{comment} We here do not impose the constraint that $R$ be
  connected as it is often done in the literature. This simplifies the
  numerical determination of the size of the largest homogeneous
  region. Results regarding the phase structure do not depend on
  whether or not connectedness is assumed in the definition of
  homogeneous regions. In particular the ordered phase consists of one
  region spanning the entire system so that both definitions yield the
  same characterisation of the low-$q$ phase.
\bibitem{pgnum} Integration of the Master equation yields a positive value for the fraction of active bonds in the ordered phase (coarsening lasts indefinitely in infinite systems in the ordered phase), and a vanishing fraction in the disordered phase, similarly to \cite{Castellano}. The location of the phase transition in Fig. \ref{fig:pg_thresh} is obtained as the point where the number of active bonds decreases below a threshold of order $10^{-3}$, or equivalently as the point at which a discontinuity in $P_F$ or its derivative occurs.
\bibitem{comment3} We here note that the data shown in Fig. \ref{fig:gammaprime} suggests that $P_F$ at stationarity might potentially be a decreasing function of $\delta$ for $\gamma\gtrsim0.01$. $P_F$ is small for all tested values of $\delta$ in this regime, however, so that we have not investigated this regime further.
\end{thebibliography}
\end{document}